\documentclass[pra,aps,amsmath,amssymb,amsfonts,twocolumn,nofootinbib,longbibliography]{revtex4-1}
\usepackage{amssymb}
\usepackage{bm,mathrsfs}
\usepackage{graphicx}
\usepackage{epsfig}
\usepackage{amsmath,bbm}
\usepackage{amsfonts,amssymb}
\usepackage{times}
\usepackage{verbatim}
\usepackage[sort&compress]{natbib}
\usepackage{amsmath}
\usepackage[colorlinks,breaklinks,linkcolor=blue,anchorcolor=blue,citecolor=blue,urlcolor=blue,dvipdfm]{hyperref}

\begin{document}

\title{Deterministic generation of maximally discordant mixed states by dissipation}

\author{X. X. Li}
\affiliation{Center for Quantum Sciences and School of Physics, Northeast Normal University, Changchun 130024,  China}
\affiliation{Center for Advanced Optoelectronic Functional Materials Research, and Key Laboratory for UV Light-Emitting Materials and Technology
of Ministry of Education, Northeast Normal University, Changchun 130024, China}

\author{H. D. Yin}
\email{yinhd239@nenu.edu.cn}
\affiliation{Center for Quantum Sciences and School of Physics, Northeast Normal University, Changchun 130024, China}
\affiliation{Center for Advanced Optoelectronic Functional Materials Research, and Key Laboratory for UV Light-Emitting Materials and Technology
of Ministry of Education, Northeast Normal University, Changchun 130024, China}

\author{D. X. Li}
\affiliation{Center for Quantum Sciences and School of Physics, Northeast Normal University, Changchun 130024, China}
\affiliation{Center for Advanced Optoelectronic Functional Materials Research, and Key Laboratory for UV Light-Emitting Materials and Technology
of Ministry of Education, Northeast Normal University, Changchun 130024, China}

\author{X. Q. Shao}
\email{shaoxq644@nenu.edu.cn}
\affiliation{Center for Quantum Sciences and School of Physics, Northeast Normal University, Changchun 130024, China}
\affiliation{Center for Advanced Optoelectronic Functional Materials Research, and Key Laboratory for UV Light-Emitting Materials and Technology
of Ministry of Education, Northeast Normal University, Changchun 130024, China}

\begin{abstract}
Entanglement can be considered as a special quantum correlation, but not the only kind. Even for a separable quantum system, it is allowed to exist non-classical correlations. Here we propose two dissipative schemes for generating a maximally correlated state of two qubits in the absence of quantum entanglement, which was raised by [F. Galve, G. L. Giorgi, and R. Zambrini, {\color{blue}Phys. Rev. A {\bf 83}, 012102 (2011)}]. These protocols take full advantages of the interaction between four-level atoms and strongly lossy optical cavities. In the first scenario, we alternatively change the phases of two classical driving fields, while the second proposal introduces a strongly lossy coupled-cavity system. Both schemes can realize all Lindblad terms required by the dissipative dynamics, guaranteeing the maximally quantum dissonant state to be the unique steady state for a certain subspace of system. Moreover, since the target state is a mixed state, the performance of our method is evaluated by the definition of super-fidelity $G(\rho_{1},\rho_{2})$, and the strictly numerical simulations indicate that fidelity outstripping $99\%$ of the quantum dissonant state is achievable with the current cavity quantum electrodynamics parameters.
\end{abstract}

\maketitle
\section{Introduction}
As one of the most striking features in quantum theory, quantum entanglement is recognized as the essential resource for quantum information processing \cite{PhysRev.47.777}. For instance, it is widely used in quantum key distribution \cite{PRA.62.044302Ref3}, superdense coding \cite{PRA.62.044302Ref2}, quantum teleportation \cite{PRA.62.044302Ref1}, and quantum computation \cite{Nature434/169/2005}. Theoretically, maximally entangled states (like Bell states) have the best performance in above tasks. However, in reality, the decoherence effect due to the environment makes the pure entangled state into a statistical mixture and degrade quantum entanglement. It is natural to ask that whether mixed state is useful for quantum information or not. The answer is positive, for example, Werner state is a typical mixed state, which is defined by a class of two-body quantum mixtures. It has many features like invariant under the unitary transformation \cite{PhysRevA.40.4277}, which has been used in the description of noisy quantum channels, such as nonadditivity claims and the study of deterministic purification \cite{AMP}.

Quantum discord, a measure of the total quantum correlations, is defined as the difference between the quantum mutual information and the classical correlations at the quantum level \cite{PhysRevLett.88.017901}. It attempts to quantify all quantum correlations including entanglement. The study of quantum discord has a crucial importance for the full development of new quantum technology because it is more robust than entanglement against the effects of decoherence \cite{arX1802.058877v1Ref4,arXRef5}. Discord between bipartite systems can be consumed to encode information with some constraints on measurement. Researchers have experimentally encoded information within the discordant correlations of two separable Gaussian states to use discord as a physical resource \cite{EPLRef8}. Especially, Glave \textit{et al.} found some mixed states have greater values of quantum discord than pure states \cite{PhysRevA.83.012102}, and they identified the family of mixed states which maximize the discord for a given value of the classical correlations. On the basis of this work, L\'{o}pez \textit{et al.} mathematically described a method to produce the maximally correlated states without entanglement \cite{EPL} and gave an example of unitary dynamic process, which restricts the evolution time of system.

The quantum dissipation characterized by a Lindblad generator in Markovian quantum master equations originates from the weak coupling between quantum systems and environment. Traditionally, it has been considered as a detrimental effect on quantum information processing. Nevertheless, appearances of various dissipative schemes show that the environment can be used as a resource for quantum computation and entanglement generation \cite{PhysRevA.59.2468,PhysRevLett.88.197901,OLRef8,OLRef9,OLRef10,OLRef11,OLRef12,OLRef13,OLRef14,PhysRevA.98.042310,
PhysRevA.98.062338,PhysRevA.99.032348,PhysRevA.90.054302,PhysRevA.92.022328,PhysRevLett.120.093601,PhysRevA.97.032328,Yang2019}. In particular, Kastoryano \textit{et al.} \cite{OLRef9} have discussed how to prepare highly entangled states via the loss of photon from an optical cavity. In Ref.~\cite{OLRef11}, the authors proposed a dissipative scheme to generate a maximally entanglement between two Rydberg atoms, where the spontaneous emissions of atoms play a positive role. Emanuele \textit{et al.} presented and analyzed a new approach for the generation of atomic spin-squeezed states using the interaction between four-level atoms and a single-mode cavity \cite{PhysRevLett.110.120402}.

Enlightened by the work of Ref.~\cite{PhysRevLett.110.120402}, we construct two physical models by taking the environment as a resource to generate the maximally discordant mixed state. This approach has following advantages: (i) Compared with the unitary dynamic evolution, the dissipative process is independent of time. (ii) The initial state is not strictly required by both schemes, and the target state can be successfully prepared as long as the state $|\Psi^{-}\rangle=(|01\rangle-|10\rangle)/\sqrt{2}$ is not populated initially. (iii) The investigated systems make full use of the cavity decay rate $\kappa$ while the spontaneous emission rates $\gamma$ of atoms are suppressed . Therefore, the parameters $\kappa$ and $\gamma$ are permitted to have a wide range of values to improve the experimental feasibility.

The remainder of the paper is organized as follows. In Sec.~\ref{II}, we briefly review the properties of maximally discordant mixed states. In Sec.~\ref{III}, we construct one physical model with a pair of four-level atoms trapped in a strongly lossy optical cavity. Under the large decay of cavity and alternatively changing the Rabi frequencies of classical fields, we derive an effective master equation and numerically simulate the effects of relevant parameters on the prepared state. In Sec.~\ref{IV}, we introduce another physical model which requires a coupled-cavity with atoms separately trapped in each cavity. In Sec.~\ref{V} and Sec.~\ref{VI} we discuss the potential experimental feasibility and give a brief summary of the work, respectively.

\section{Brief Review of the maximally discordant mixed states}\label{II}
The states we are interested in are found within the set of separable states. It has been shown that the most nonclassical two-qubit states, {i.e.}, the family with maximal quantum discord versus classical correlations, were formed by mixed states of rank 2 and 3, which are named maximally discordant mixed states (MDMS). The class of states of rank 3 is defined by \cite{PhysRevA.83.012102}
\begin{equation}
\rho=\epsilon|\Phi^{+}\rangle\langle\Phi^{+}|+(1-\epsilon)[x|01\rangle\langle01|+(1-x)|10\rangle\langle10|],
\end{equation}
where $|\Phi^{+}\rangle=(|00\rangle+|11\rangle)/\sqrt{2}$.

Quantum discord is defined as $I-C$, where $I=S(\rho_{A})+S(\rho_{B})-S(\rho_{AB})$ is quantum mutual information, of which $S(\rho)$ is Von neumann entrophy and $C(\rho_{AB})={\bf max}\{S(\rho_{A})-S(\rho_{A|B})\}$ is classical correlation where $S(\rho_{A|B})$ is the conditional entropy of $A$ given a measurement on the system $B$. Refer to Ali-Rau-Alber results of the conditional entropy \cite{PRA.81.042105}, the quantum mutual information is maximized when $x=1/2$ and $\epsilon=1/3$, meanwhile the classical correlation is minimized, which corresponds to a maximally discordant mixed state. By exchanging the basis vector $|0\rangle\leftrightarrow|1\rangle$ of the second qubit, we obtain the state in the form:
\begin{equation}
\rho=\frac{1}{3}(|\Psi^{+}\rangle\langle\Psi^{+}|+|00\rangle\langle00|+|11\rangle\langle11|),
\end{equation}
where $|\Psi^{+}\rangle=(|01\rangle+|10\rangle)/\sqrt{2}$.

Using the basis of Bell states $|\Phi^{\pm}\rangle=(|00\rangle\pm|11\rangle)/\sqrt{2}$ and $|\Psi^{\pm}\rangle=(|01\rangle\pm|10\rangle)/\sqrt{2}$~\cite{EPL}, the above state can be rewritten as:
\begin{equation}\label{003}
\rho=\frac{1}{3}(|\Phi^{+}\rangle\langle\Phi^{+}|+|\Phi^{-}\rangle\langle\Phi^{-}|+|\Psi^{+}\rangle\langle\Psi^{+}|).
\end{equation}
When we have a system characterized by the following master equation
\begin{equation}\label{004}
\dot{\rho}=\mathscr{L}_{\gamma_{x}}[S_{x}]\rho+\mathscr{L}_{\gamma_{y}}[S_{y}]\rho,
\end{equation}
where $S_{x}=(\sigma_{x}^{1}+\sigma_{x}^{2})$ and $S_{y}=(\sigma_{y}^{1}+\sigma_{y}^{2})$ ($\sigma_{x,y}$ are pauli operators), $\mathscr{L}_{\gamma_{i}}[O]\rho$ is the Lindblad term  defined as $\mathscr{L}_{\gamma_{i}}[O]\rho=\gamma_{i}/2(2O\rho O^{\dag}-O^{\dag}O\rho-\rho O^{\dag}O), (i=x,y)$, the state described by Eq.~(\ref{003}) will be the steady state of this system. However, it is difficult to find a natural system with the  above form of the master equation. Thence we consider to design a physical model which is equivalent to Eq.~(\ref{004}) under the appropriate approximations, and we will discuss our method detailedly in the next section.

\section{two four-level atoms in a lossy cavity}\label{III}
The central idea of our work can be understood by considering a pair of atoms interacting with a  strongly lossy optical cavity, as depicted in Fig.~\ref{p1}. The atoms are driven by the laser fields with complex Rabi frequencies $\Omega_{1(2)}e^{i\varphi_{1(2)}}$, where $\varphi_{1(2)}$ is the phase of classical field, and simultaneously coupled to the quantized field with strength $g$. The Hamiltonian in the Schr\"{o}dinger picture can be written as ($\hbar=1$):
\begin{figure}
\centering\scalebox{0.25}{\includegraphics{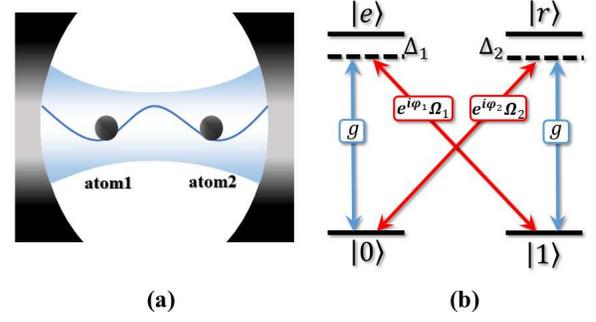}}
\caption{\label{p1}Schematic view of the system and the configuration of the atoms. (a) The system consists of two atoms collectively interacting with a lossy cavity. (b) Level structure of a four-level atom which is simultaneously driven by two classical fields and coupled to a cavity mode.}
\end{figure}
\begin{eqnarray}
H_{s}&=&H_{0}+V_{s},\\
H_{0}&=&\sum_{i=1}^{2}\omega_{0}|0\rangle_{i}\langle0|+\omega_{1}|1\rangle_{i}\langle1|+\omega_{e}|e\rangle_{i}\langle e|\nonumber\\&&+\omega_{r}|r\rangle_{i}\langle r|+\nu a^{\dag}a,\nonumber\\
V_{s}&=&\sum_{i=1}^{2}g(|e\rangle_{i}\langle0|+|r\rangle_{i}\langle1|)a+\Omega_{1}e^{i\varphi_{1}}|e\rangle_{i}\langle1|e^{-i\mu_{1}
t}\nonumber\\&&+\Omega_{2}e^{i\varphi_{2}}|r\rangle_{i}\langle0|e^{-i\mu_{2}t}+{\rm H.c.},\nonumber
\end{eqnarray}
where $\omega_{0}$, $\omega_{1}$, $\omega_{e}$, and $\omega_{r}$ are the eigenfrequencies of the lower states $|0\rangle$, $|1\rangle$ and upper states $|e\rangle$, $|r\rangle$, respectively, while $\nu$ and $\mu_{1(2)}$ are the frequencies of quantum and classical fields. $a^{\dag}$ and $a$ are the creation and annihilation operators of the optical cavity mode. In addition, the ground states transition is dipole-forbidden.
For simplicity, we assume all parameters are real. In the interaction picture, the Hamiltonian of the system reads:
\begin{eqnarray}\label{006}
H_I&=&H_{1}+H_{2},\\
H_{1}&=&\sum_{i=1}^{2}ga|e\rangle_{i}\langle 0|e^{i\delta_{1} t}+\Omega_{1}e^{i\varphi_{1}}|e\rangle_{i}\langle 1|e^{i\delta_{2} t}+{\rm H.c.},\nonumber\\
H_{2}&=&\sum_{i=1}^{2}ga|r\rangle_{i}\langle 1|e^{i\delta'_{1} t}+\Omega_{2}e^{i\varphi_{2}}|r\rangle_{i}\langle 0|e^{i\delta'_{2} t}+{\rm H.c.},\nonumber
\end{eqnarray}
where $\delta_{1(2)}=\omega_{e}-\omega_{0(1)}-\nu(\mu_{1})$ and $\delta'_{1(2)}=\omega_{r}-\omega_{1(0)}-\nu(\mu_{2})$.
We further suppose $\delta_{1}=\delta_{2}=\Delta_{1}$ and $\delta'_{1}=\delta'_{2}=\Delta_{2}$.
Now we consider the process of constructing the collective decay operator $S_{y}=\sigma_{y}^{1}+\sigma_{y}^{2}$. Taking $ \Omega_{1}e^{i\varphi_{1}}=i\Omega_{1}$, and $\Omega_{2}e^{i\varphi_{2}}=-i\Omega_{2}$, and in the regime of large detuning $|\Delta_{1(2)}|\gg\{g,\Omega_{1(2)}\}$, we can safely eliminate the upper states $|e\rangle$ and $|r\rangle$, then the above Hamiltonian reduces to
\begin{eqnarray}
H_{\textmd{eff}}&=&H_{\textmd{eff}}^1+H_{\textmd{eff}}^2,
\end{eqnarray}
where
\begin{eqnarray}
H_{\textmd{eff}}^1&=&G_{1}J_{-}a^{\dag}+{\rm H.c.}+\sum_{i=1}^{2}g_{\textmd{eff}}^{1}a^{\dag}a|0\rangle_{i}\langle0|+\Omega_{\textmd{eff}}^{1}|1\rangle_{i}\langle1|,\nonumber\\
H_{\textmd{eff}}^2&=&G_{2}J_{+}a^{\dag}+{\rm H.c.}+\sum_{i=1}^{2}g_{\textmd{eff}}^{2}a^{\dag}a|1\rangle_{i}\langle1|+\Omega_{\textmd{eff}}^{2}|0\rangle_{i}\langle0|,\nonumber
\end{eqnarray}
with $G_{1(2)}=g\Omega_{1(2)}/\Delta_{1(2)}$, $g_{\textmd{eff}}^{1(2)}=-g^{2}/\Delta_{1(2)}$, and $\Omega_{\textmd{eff}}^{1(2)}=-\Omega_{1(2)}^{2}/\Delta_{1(2)}$. $J_{+}=i(|1\rangle_{1}\langle 0|+|1\rangle_{2}\langle 0|)$ and $J_{-}=-i(|0\rangle_{1}\langle 1|+|0\rangle_{2}\langle 1|)$ are the collective ascending and descending operators. We further assume $G_{1}=G_{2}=G$, {i.e.}, $\Omega_{1}/\Delta_{1}=\Omega_{2}/\Delta_{2}$, and omit the Stark shifts of the ground states, the above Hamiltonian is simplified as
\begin{equation}
H_{\textmd{eff}}=G(J_{-}+J_{+})a^{\dag}+{\rm H.c.}.
\end{equation}
Since the effective system only includes the ground states, the spontaneous emissions of atoms are greatly restrained, and  the master equation could be written as
\begin{eqnarray}\label{master}
\dot{\rho}&=&-i[G(J_{-}+J_{+})a^{\dag}+G(J_{-}+J_{+})a,\rho]\nonumber\\&&+\frac{\kappa}{2}(2 {a}\rho {a}^{\dag}- {a}^{\dag} {a}\rho-\rho {a}^{\dag} {a}).
\end{eqnarray}
In the limitation of large decay rate $\kappa\gg{G}$, the cavity mode can also be neglected, and we obtain the master equation characterizing the system of atoms as:
\begin{equation}\label{mastery}
\dot{\rho}=\mathscr{L}_{\gamma_{y}}[S_{y}]\rho,
\end{equation}
where $\gamma_{y}=4G^{2}/\kappa$ is the collective decay rate of the atoms.
\begin{figure}
\centering\scalebox{0.38}{\includegraphics{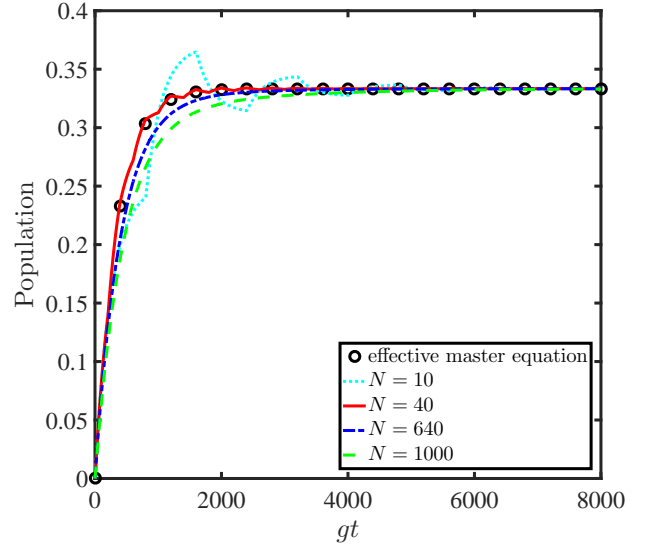}}
\caption{\label{p2}The populations of $|\Psi^{+}\rangle$ as functions of $gt$ governed by the effective master equation (\ref{Leff}) and the master equations with Hamiltonian (\ref{006}), where $N$ is the switching number. The initial state is $|00\rangle|0\rangle_{c}$ and we set $G=0.01$ and $\kappa=80G$.}
\end{figure}

On the other hand, if we attempt to construct the collective decay operator $S_{x}=\sigma_{x}^{1}+\sigma_{x}^{2}$, we can simply take $\varphi_{1}=\varphi_{2}=0$, then after a series of similar derivations, and the effective master equation reads
\begin{equation}\label{masterx}
\dot{\rho}=\mathscr{L}_{\gamma_{x}}[S_{x}]\rho,
\end{equation}
where $\gamma_{x}=4G^{2}/\kappa$, $J_{+}^{'}=|1\rangle_{1}\langle0|+|1\rangle_{2}\langle0|$ and $J_{-}^{'}=|0\rangle_{1}\langle1|+|0\rangle_{2}\langle1|$.

Up to present, we have shown how to generate the collective decay operators $S_x$ and $S_y$ respectively. But the stability of Eq.~(\ref{003}) requires there should be $\mathscr{L}_{\gamma_{x}}(S_{x})$ and $\mathscr{L}_{\gamma_{y}}(S_{y})$  in the master equation at the same time. Fortunately, drawing lessons from the spin echoes effect, our model is able to simulate the effective master equation of  Eq.~(\ref{004}) apart from a coefficient $1/2$, as long as the phases of the classical fields $\varphi_1$ and $\varphi_2$  are interchanged fast enough. The result is obtained by using the Trotter product formula (see Corollary $5.8$ in Chap. III of Ref.~\cite{PRA201592062114Ref21})
\begin{equation}
\lim_{N\rightarrow\infty}\{e^{\mathscr{L}_{\gamma_{x}}[S_{x}]\frac{T}{2N}}e^{\mathscr{L}_{\gamma_{y}}[S_{y}]\frac{T}{2N}}\}^N
=e^{\frac{1}{2}\{\mathscr{L}_{\gamma_{x}}[S_{x}]+\mathscr{L}_{\gamma_{y}}[S_{y}]\}T},
\end{equation}
where $T$ is the total evolution time. Then the effective master equation is
\begin{equation}\label{Leff}
\dot{\rho}=\frac{1}{2}\mathscr{L}_{\gamma_{x}}[S_{x}]\rho+\frac{1}{2}\mathscr{L}_{\gamma_{y}}[S_{y}]\rho.
\end{equation}
\begin{figure}
\centering\scalebox{0.36}{\includegraphics{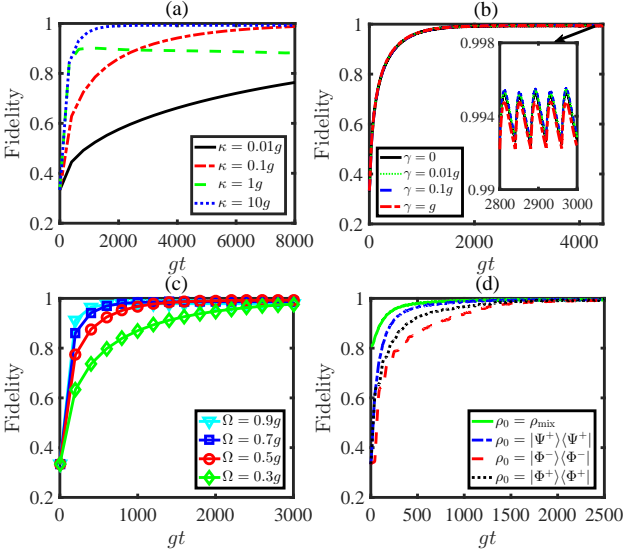}}
\caption{\label{p3}The target state fidelities as functions of $gt$ governed by the full master equation and the switching number $N=200$. (a) Time evolutions with different cavity decay rates $\kappa=(0.01, 0.1, 1, 10)g$, where we set $\Delta_{1,2}=100g$, and $\Omega_{1(2)}=0.5g$. (b) Discussion of the effects of atom spontaneous emission rates $\gamma$ with the same parameters as (a) and $\kappa=0.1g$. The inset shows the enlarge view of the part indicated by the arrow. (c) The effects of Rabi frequencies on the target state with $\gamma=0.1g$ and $\kappa=0.1g$. (d) Time evolutions with different initial states, where $|\Phi^{\pm}\rangle=(|00\rangle|0\rangle_{c}\pm|11\rangle|0\rangle_{c})/\sqrt{2}$, $|\Psi^{+}\rangle=(|01\rangle|0\rangle_{c}+|10\rangle|0\rangle_{c})/\sqrt{2}$, and $\rho_{\textmd{mix}}=(0.1|\Phi^{+}\rangle\langle\Phi^{+}|+0.1|\Phi^{-}\rangle\langle\Phi^{-}|+0.8|\Psi^{+}\rangle
\langle\Psi^{+}|)\otimes|0\rangle_{c}\langle 0|$. }
\end{figure}

Fig.~\ref{p2} shows the population of $|\Psi^{+}\rangle$ under different evolution processes from the initial state $|00\rangle|0\rangle_{c}$. The evolution of the effective master equation (\ref{Leff}) is shown with empty circles and the other lines are the switching evolutions obtained from the master equation with Hamiltonian (\ref{006}). The total evolution time is $gt=8000$. Different lines correspond to the results with different switching number $N$. Since we take the cavity decay as a resource for states generation, the switching number $N$ has an upper limit promising the interval time much larger than $1/\kappa$. This can guarantee the role of $\kappa$ in each process, which insure the complete generation of the target state. In addition, the operation time $1/\gamma$ determines the minimum value of $N$. This ensures the interval time far less than $1/\gamma$. Thus  we choose $N=200$ and $gt=8000$ in the following simulations if there is no special description.

In quantum information theory, distinguishing two quantum states is a fundamental task. One of the main tools used in distinguishability theory is quantum fidelity \cite{informationbook,arXiv.0811.3453v4} which is widely used and has been found applications in solving some problems like quantifying entanglement \cite{arXiv.0811.3453v4Ref3,arXiv.0811.3453v4Ref4}, quantum error correction \cite{PhysRevLett.100.020502}, quantum chaos \cite{PhysRevE.81.017203} and so on.
In order to measure the distance between quantum states including mixed states, we here adopt the definition of super-fidelity \cite{DBLP:journals/qic/MiszczakPHUZ09}
\begin{eqnarray}
G(\rho,\sigma)&=&\textmd{Tr}[\rho(t)\sigma]+\sqrt{[1-\textmd{Tr}\rho(t)^{2}](1-\textmd{Tr}\sigma^{2})},
\end{eqnarray}
with $\sigma$ being the density operator of the target state as $\sigma=(|\Psi^{+}\rangle\langle\Psi^{+}|+|00\rangle\langle00|+|11\rangle\langle11|)/3$. We initialize the system into state $|00\rangle|0\rangle_{c}$ and plot the fidelity of the target-state under the switching evolution of the master equations with full Hamiltonian (\ref{006}). Figure \ref{p3}(a), \ref{p3}(b), and \ref{p3}(c) respectively discuss the effects of parameters $\kappa, \gamma$, and $\Omega$ on the preparation of the target state. Fig.~\ref{p3}(a) shows the fidelity as a function of the cavity decay rate $\kappa$ with parameters $\Delta_{1,2}=100g$, and $\Omega_{1,2}=0.5g$. The increase of $\kappa$ will prolong the convergence time. It can be explained by Eqs.~(\ref{master}) and (\ref{mastery}). To obtain the target state, the collective decay rate $4G^{2}/\kappa$ will increase as $\kappa$ decreasing, which results in a short convergence time. But if $\kappa$ is too small, it will destroy the condition $\kappa\gg G$ and fail to generate the target state.

In Fig.~\ref{p3}(b), we take into account the spontaneous emissions of the atoms and plot the evolutions of the target state with different $\gamma$. Even if $\gamma$ is extremely large $(\gamma\sim g)$, the fidelity is still above $99\%$, which demonstrates that our scheme has favorable resistance to atomic spontaneous emission. The inset picture of Fig.~\ref{p3}(b) is the enlarge view of the part indicated by the arrow, which shows that the population keeps oscillating at the final time with small amplitude, and stays around a definite value.

\begin{figure}
\centering\scalebox{0.38}{\includegraphics{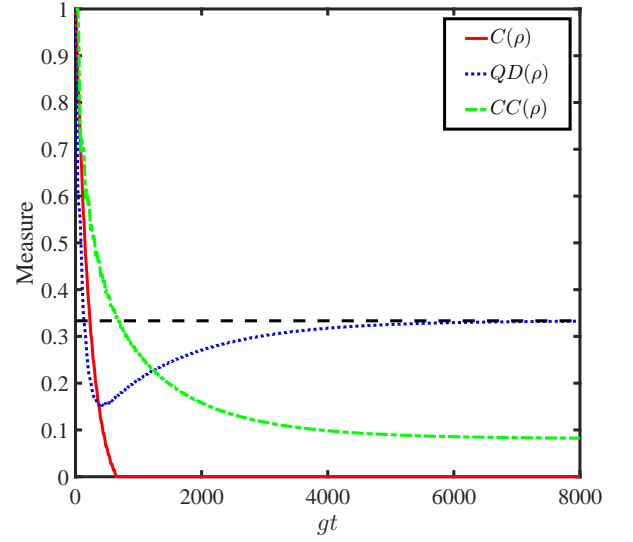}}
\caption{\label{p4}The properties of the target state, which consists of the concurrence (\emph{C}), classical correlation (\emph{CC}) and quantum discord (\emph{QD}). The initial state is $|\Psi^{+}\rangle$ and the black dashed line labels the number $1/3$.}
\end{figure}
\begin{figure}
\centering\scalebox{0.38}{\includegraphics{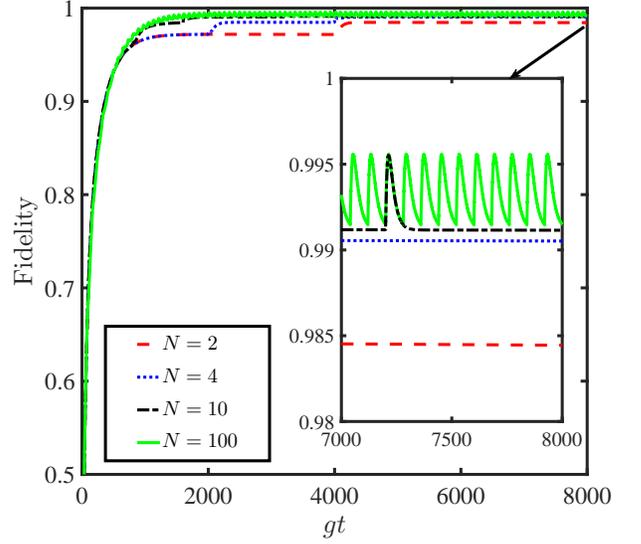}}
\caption{\label{p5}Time evolution of the fidelities for the target state with different switching numbers. The initial state is $|00\rangle|0\rangle_{c}$. The inset shows the zoom-in fidelities from $gt=7000$ to $8000$.}
\end{figure}
Moreover, the convergence time is related to the intensity of the classical field $\Omega_{1(2)}$. Fig.~\ref{p3}(c) displays the evolution curves under different $\Omega$ with $\gamma=0.1g$ and discusses the optimal parameter range of Rabi frequency. The figure shows the optimal range of $\Omega$ is about $0.3g\sim0.7g$, which could ensure the fidelity over $99\%$. Fig.~\ref{p3}(d) additionally considers the request to the initial state of the system. We can obtain the target state with arbitrary initial state except for the singlet state $|\Psi^{-}\rangle$.

To expound the properties peculiar to the target state, we plot the concurrence \cite{EofCon}, classical correlation and quantum discord of the state with the full master equation in Fig.~\ref{p4}. It is worth mentioning that we directly utilize the results given in Ref.~\cite{PLA.377.53.2012} to measure quantum discord (\emph{QD}), and the calculation of $S(\rho_{A|B})$ is based on the positive-operator-valued measurements (POVM) locally performed on the subsystem B. The \emph{QD} and the classical correlation (\emph{CC}) are given as: $QD(\rho)={\bf min}\{Q_{1},Q_{2}\}$, $CC(\rho)={\bf max}\{CC_{1},CC_{2}\}$, where $CC_{j}=h[\rho_{11}+\rho_{22}]-D_{j}$ and $Q_{j}=h[\rho_{11}+\rho_{33}]+\sum_{k=1}^{4}\lambda_{k}{\rm log}_{2}\lambda_{k}+D_{j}$, with $\lambda_{k}$ being the eigenvalues of $\rho$ and $h[x]$ is the binary entropy defined as $h[x]=-x{\rm log}_{2}x-(1-x){\rm log}_{2}(1-x)$. $D_{1}=h[\tau]$, where $\tau=(1+\sqrt{[1-2(\rho_{33}+\rho_{44})]^{2}+4(|\rho_{14}|+|\rho_{23}|)^{2}})/2$ and $D_{2}=-\sum_{k=1}^{4}\rho_{kk}{\rm log}_{2}\rho_{kk}-h[\rho_{11}+\rho_{33}]$. Based on Fig.~\ref{p4}, the final state has the maximally quantum discord $1/3$ without entanglement and the classical relation reaches the minimum. The steady state is a maximally discordant mixed state.

Fig.~\ref{p5} discusses the effect of the switching number $N$, where the increasing of $N$ smooths the evolution process. It also illustrates that a high fidelity over $98\%$ can be obtained with a wide range of values for $N$. Even if $N=4$ the fidelity can still get over $99\%$. Thus, in actual operations, we can properly reduce the value of $N$ to simplify the experiment.

\section{two atoms in a lossy coupled-cavity system}\label{IV}
\begin{figure}
\centering\scalebox{0.19}{\includegraphics{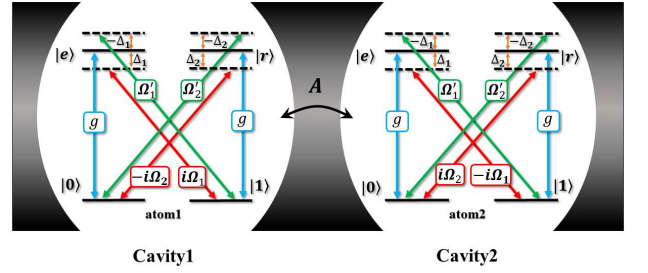}}
\caption{\label{p6}Schematic view of two four-level atoms trapped in a lossy coupled-cavity array. Each atom is simultaneously driven by four classical fields with Rabi frequencies $\pm i\Omega_{1(2)}, \Omega'_{1(2)}$, detuned by $\pm\Delta_{1(2)}$ and resonantly coupled with the corresponding cavity mode. The photon can hop between two cavities with coupling strength $A$.}
\end{figure}

The coupled-cavity systems are especially useful in distributed quantum computation \cite{PhysRevA.77.033801,PhysRevLett.101.246809,Liew_2013,Laser,PhysRevLett.96.010503}, which are able to overcome the problem
of individual addressability.
In our model, the lossy coupled-cavity system  is shown in Fig.~\ref{p6}. It consists of two coupled cavities which respectively trapped a four-level atom with ground states $|0\rangle, |1\rangle$ and excited states $|e\rangle, |r\rangle$. The transition between $|0\rangle$ ($|1\rangle$) and $|e\rangle$ ($|r\rangle$) is coupled resonantly to the quantum field with coupling constant $g$, and other non-resonant transitions with detuning $\pm\Delta$ are driven by classical fields with Rabi frequencies $\pm i\Omega_{1(2)}$ and $\Omega'_{1(2)}$. The Hamiltonian under the Schr\"{o}dinger picture can be written as
\begin{eqnarray}\label{e1}
H&=&H_0+V_s,\\
H_0&=&\sum_{k=1}^2\omega_0|0\rangle_k\langle 0|+\omega_1|1\rangle_k\langle 1|+\omega_e|e\rangle_k\langle e|\nonumber\\&&+\omega_r|r\rangle_k\langle r|+\omega_c a_k^\dag a_k,\nonumber\\
V_s&=&\sum_{k=1}^2g(|e\rangle_{k}\langle0|+|r\rangle_{k}\langle1|)a_{k}+\Omega_{1}'|e\rangle_{k}\langle1|e^{-i\omega_L't}
\nonumber\\&&+\Omega_{2}'|r\rangle_{k}\langle0|e^{-i\omega_L't}
+i\Omega_{1}(|e\rangle_{1}\langle 1|-|e\rangle_{2}\langle 1|)e^{-i\omega_{L}t}
\nonumber\\&&-i\Omega_{2}(|r\rangle_{1}\langle0|-|r\rangle_{2}\langle0|)e^{-i\omega_{L}t}+\mathrm{H.c.}\nonumber\\&&+A(a_{1}^\dag a_{2}+a_{2}^\dag a_{1}),\nonumber
\end{eqnarray}
where $\omega_i$ $(i=0,1,e,r)$ are the eigenfrequencies of ground and excited states for each atom, $\omega_c$ is the frequency of quantum field. $a_k^\dag$ and $a_k~(k=1,2)$ are creation and annihilation operators of cavity mode $k$, $\omega_L$ and $\omega_L'$ are frequencies of classical fields. We switch the Hamiltonian from Schr\"{o}dinger picture to the interaction picture and obtain
\begin{eqnarray}\label{e5}
H_I&=&\sum_{k=1}^2g(|e\rangle_k\langle 0|+|r\rangle_k\langle 1|)a_k+i\Omega_{1} e^{i\Delta_{1} t}(-1)^{k-1}|e\rangle_{k}\langle 1|\nonumber\\&&+i\Omega_{2} e^{i\Delta_{2} t}(-1)^{k}|r\rangle_{k}\langle 0|+\Omega_{1}' e^{-i\Delta_{1} t}|e\rangle_k\langle 1|\nonumber\\&&+\Omega_{2}' e^{-i\Delta_{2} t}|r\rangle_k\langle 0|+\mathrm{H.c.}+A(a_1^\dag a_2+a_2^\dag a_1),
\end{eqnarray}
where $\Delta_{1}=\omega_e-\omega_1-\omega_L=\omega_1+\omega_L'-\omega_e$,
$\Delta_{2}=\omega_r-\omega_0-\omega_L=\omega_0+\omega_L'-\omega_r$, and we suppose $\Delta_{1}=\Delta_{2}=\Delta$.
Now we introduce a pair of delocalized bosonic modes in order to remove the localized modes as follows \cite{PhysRevLett.96.010503},
\begin{eqnarray}\label{e6}
m_1\equiv\frac{1}{\sqrt{2}}(a_1-a_2),\  \  m_2\equiv\frac{1}{\sqrt{2}}(a_1+a_2).
\end{eqnarray}
Then we have
\begin{eqnarray}\label{e8}
H_I&=&\sum_{k=1}^2\frac{g}{\sqrt{2}}m_1e^{iAt}(-1)^{k-1}(|e\rangle_k\langle 0|+|r\rangle_k\langle 1|)\nonumber\\&&+\frac{g}{\sqrt{2}}m_2e^{-iAt}(|e\rangle_k\langle 0|+|r\rangle_k\langle 1|)+\Omega'_{1} e^{-i\Delta t}|e\rangle_k\langle 1|\nonumber\\
&&+\Omega'_{2} e^{-i\Delta t}|r\rangle_k\langle 0|+i\Omega_{1} e^{i\Delta t}(-1)^{k-1}|e\rangle_{k}\langle 1|\nonumber\\&&+i\Omega_{2} e^{i\Delta t}(-1)^{k}|r\rangle_{k}\langle 0|+\mathrm{H.c.}.
\end{eqnarray}
We set $A=\Delta$ to guarantee the two-photon resonance, and choose $\Omega_{1(2)}=\Omega'_{1(2)}=\Omega$. Under the large detuning condition, \emph{i.e.}, $|\Delta|\gg\left\{g, \Omega\right\}$, and neglecting the Stark-shift terms, the effective Hamiltonian reads
\begin{eqnarray}\label{e9}
H_{{\textmd{eff}}}&=&\sum_{k=1}^2\frac{g\Omega}{\sqrt{2}\Delta}m_1(-i|0\rangle_k\langle 1|+i|1\rangle_{k}\langle 0|)\nonumber\\&&+\frac{g\Omega}{\sqrt{2}\Delta}m_2(|0\rangle_k\langle 1|+|1\rangle_k\langle 0|)+\mathrm{H.c.}.
\end{eqnarray}
Based on the definition of collective decay operators $S_{x}=(\sigma_{1x}+\sigma_{2x}), S_{y}=(\sigma_{1y}+\sigma_{2y})$, the effective Hamiltonian can be rewritten as
\begin{equation}\label{e10}
H_{\textmd{eff}}=Gm_1S_y+Gm_2S_x+\mathrm{H.c.},
\end{equation}
where $G=g\Omega/\sqrt{2}\Delta$. It can be seen that the current system only involves couplings between ground states and  delocalized cavity modes.
\begin{figure}
\centering\scalebox{0.38}{\includegraphics{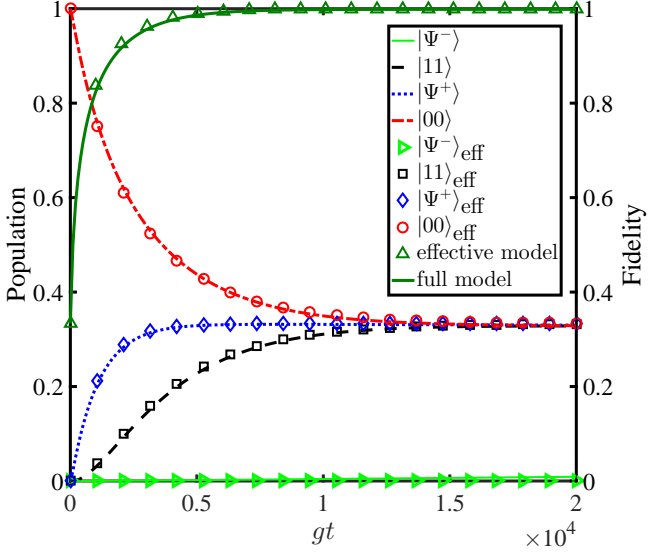}}
\caption{\label{p7}Time evolution of populations (shown in the red dash-dotted line, blue dotted line, black dashed line and green full line) and fidelities (shown in dark green) of the target state under the full and effective master equations, where $\kappa=0.1g, \Omega=0.2g, \Delta=100g$, and $\gamma=0$.}
\end{figure}
\begin{figure}
\centering\scalebox{0.36}{\includegraphics{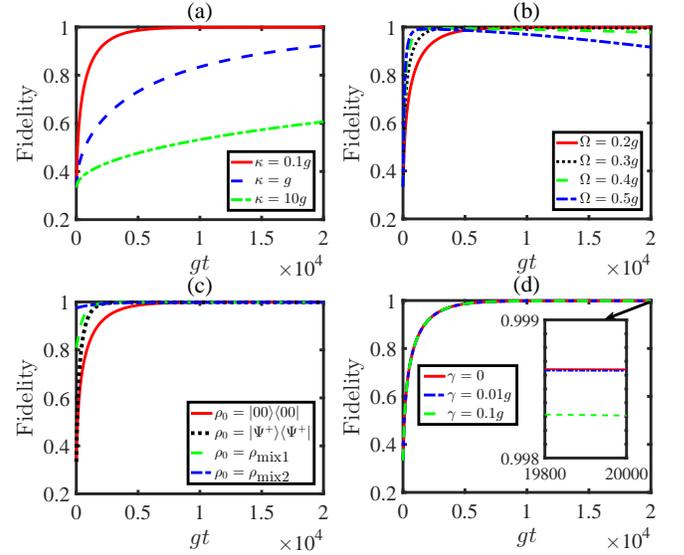}}
\caption{\label{p8}Time evolutions of the target state fidelities under the full master equation with different parameters. (a) Different curves correspond to different cavity decay rates $\kappa$. The other parameters are $\Delta=100g$, and $\Omega=0.2g$. (b) The effects of Rabi frequencies $\Omega$ on the target state with $\kappa=0.1g$ and the other parameters are same as (a). (c) Analyse different evolution processes under different initial states, where $|\Psi^{+}\rangle=(|01\rangle|00\rangle_{c}+|10\rangle|00\rangle_{c})/\sqrt{2}$, $\rho_{\textmd{mix1}}=(0.1|00\rangle\langle00|+0.1|11\rangle\langle11|+0.8|\Psi^{+}\rangle
\langle\Psi^{+}|)\otimes|00\rangle_{c}\langle 00|$, and $\rho_{\textmd{mix2}}=(0.2|00\rangle\langle00|+0.5|11\rangle\langle11|+0.3|\Psi^{+}\rangle
\langle\Psi^{+}|)\otimes|00\rangle_{c}\langle 00|$. (d) Discussion of the atom spontaneous emission rates $\gamma=(0, 0.01, 0.1)g$. The other parameters are same as above. The inset picture is the enlarge view of the part indicated by the arrow.}
\end{figure}
Therefore, the dissipative dynamics of system can be considered as governed by the following master equation
\begin{equation}\label{e10}
\dot{\rho}=i[\rho,H_{\textmd{eff}}]+\sum_{k=1}^2\frac{\kappa}{2}(2m_k \rho m_k^\dag-m_k^\dag m_k \rho-\rho m_k^\dag m_k).
\end{equation}
In the limit $\kappa\gg |G|$, we can adiabatically eliminating the delocalized cavity modes, and obtain the effective master equation,
\begin{equation}\label{e12}
\dot{\rho}=\mathscr{L}_{\gamma_{x}}[S_{x}]\rho+\mathscr{L}_{\gamma_{y}}[S_{y}]\rho,
\end{equation}
where $\gamma_{x}=\gamma_{y}=4G^{2}/\kappa$. Compared with the previous model, the coupled-cavity system provides the mean to realize $\mathscr{L}_{\gamma_{x}}[S_{x}]$ and $\mathscr{L}_{\gamma_{y}}[S_{y}]$ simultaneously. Thus the target state $\rho=(|\Psi^+\rangle\langle\Psi^+|+|00\rangle\langle 00|+|11\rangle\langle 11|)/3$ can be generated using the driven-dissipative dynamics.

To verify the effectiveness of our scheme in generating MDMS, we respectively plot the population and the fidelity of the target state with the initial state $|00\rangle|00\rangle_{c}$ governed by the full and effective master equation in Fig.~\ref{p7}. We can find that these two lines perfectly coincide with each other and the state prepared by our scheme can maintain a high fidelity close to unity after $gt=10000$. The selections of numerical simulation parameters are $\kappa=0.1g$, $\Omega=0.2g$, and  $\Delta=100g$.

Then we make the same discussions as Fig.~\ref{p3} in the coupled-cavity system. The results are shown in Fig.~\ref{p8}, which show similar phenomena of $\kappa, \rho_{0}, \gamma$ and $\Omega$. Compared with the first scenario, the fidelity is higher and the final population is stable after a longer evolution time.

\section{Discussion}\label{V}
\begin{figure}
\centering\scalebox{0.20}{\includegraphics{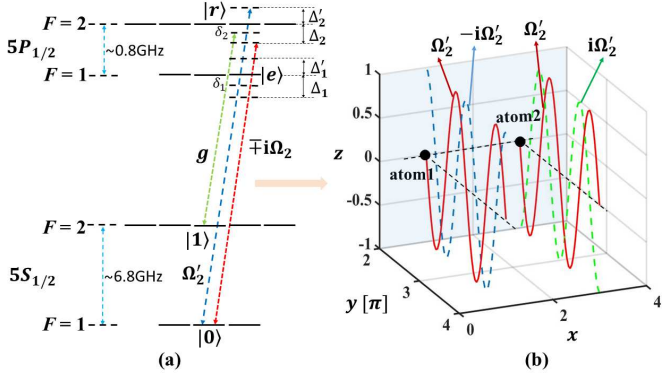}}
\caption{\label{p10}Experimental level-scheme in $^{87}\textmd{Rb}$ using the clock state of the second scenario. Together shows a set of phase relations mach the conditions.}
\end{figure}

Now, we discuss about the basic elements that maybe candidate for the intended experiment. The possible realizations of these physical models could be set up in $^{87}\textmd{Rb}$ using the clock states $|F=1,m_{F}=0\rangle$ and $|F=2,m_{F}=0\rangle$ in the $5S_{1/2}$ ground-state manifold as two-lower levels $|0\rangle$ and $|1\rangle$. In addition, the states $|F=1,m_{F}=+1\rangle$, and $|F=2,m_{F}=+1\rangle$ of the $5P_{1/2}$ manifold as two-higher levels $|e\rangle$ and $|r\rangle$ \cite{PhysRevLett.110.120402}. The possible realizations of these physical models are indicated in Fig.~\ref{p10}(a),
in which we only show the couplings between two polarization-dependent lasers and the four-level atoms in the coupled-cavity system for simplicity.
{Fig.~\ref{p10}(b) provides a method of the alignment of lasers. We use two pulses traveled in $y$ direction driving two atoms in $xz$ plane respectively. Since the Rabi frequency is presented as $\Omega_{1(2)}e^{i\varphi_{1(2)}}$ which cannot be simply displayed, we only plot the imaginary part to illustrate the phase relations. Here we take $xz$ as a reference plane (shadow area) and $\Omega'_{2}$ as a standard pulse whose phase equals to zero ($\varphi_{2}=0$). The other pulses $\pm i\Omega_{2}$ can be obtained by modulating the initial phases $\varphi_{2}=\pm\pi/2$ relative to the standard classical field}. Thus, we could construct a group of pulses to meet the above phase conditions. Note that we only show the coupling between $|0\rangle$ and $|r\rangle$ here, while to the coupling between $|1\rangle$ and $|e\rangle$, we can do similar operations.

According to past works \cite{nature06120,PhysRevA.82.053832,PhysRevLett.110.090402,NJP.16.043020.2014}, the transition between the atomic ground level $5S_{1/2}$ and the optical level $5P_{1/2}$ of $^{87}\textmd{Rb}$ atom is coupled to the quantized cavity mode with strength $g=2\pi\times14.4~\textmd{MHz}$. The spontaneous emission rate is $\gamma=2\pi\times3~\textmd{MHz}$ and the cavity decay rate is $\kappa=2\pi\times0.66~\textmd{MHz}$. The Rabi frequencies $\Omega_{1,2}$ can be tuned continuously and for the first scheme we adopt parameters $\Omega_{1,2}=0.3g$, $\Delta_{1,2}=76g$, and $N=200$, the fidelity of the target state is $99.41\%$. For the second one, we set $\Omega_{1,2}=\Omega'_{1,2}=0.1g, \Delta_{1,2}=50g$ and the fidelity is $99.56\%$.

In addition, Ref.~\cite{PhysRevA.71.013817} reported the projected limits for a Fabry-P\'{e}rot cavity, where the coupling coefficient $g=2\pi\times770$~{MHz}. Based on the corresponding critical photon number and critical atom number, we obtain $(\kappa,\gamma)=2\pi\times(21.7,2.6)~\textmd{MHz}$. The fidelity $F$ reaches $99.10\%$ for the first scheme with the other relevant parameters are selected as $\Omega_{1,2}=0.2g$, $\Delta_{1,2}=72g$, and $N=200$. For the second one, the fidelity is $99.67\%$, while other parameters are $\Omega_{1,2}=\Omega'_{1,2}=0.12g$ and $\Delta_{1,2}=50g$. Moreover, in a microscopic optical resonator \cite{Science.319.1062.2008}, the parameters of an atom interacting with an evanescent field are $(g,\kappa,\gamma)=2\pi\times(70,5,1)~\textmd{MHz}$, which correspond to the fidelity $F=99.18\%$ with parameters $\Omega_{1,2}=0.3g, \Delta_{1,2}=43g, N=200$ in the first scheme and $F=99.34\%$ with parameters $\Omega_{1,2}=\Omega'_{1,2}=0.1g, \Delta_{1,2}=50g$ in the second scheme.

\begin{figure}
\centering\scalebox{0.36}{\includegraphics{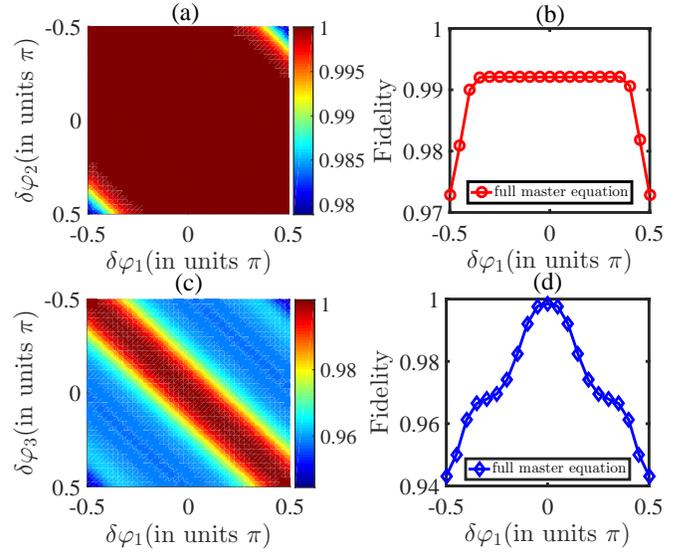}}
\caption{\label{p9}Phase discussion of aforementioned systems. (a) (b) respectively demonstrate how phase mismatch affect the first scheme with effective and full master equations, while (c) (d) show the phase mismatch influences to the second one also with effective and full master equations,respectively. The parameters are taken as (a) $\gamma t=3$ (b) $\Omega=0.2g$, $\Delta=100g$, $N=40$, and $gt=12000$ (c) $\gamma t=8$ (d) $\Omega=0.5g$, $\Delta=100g$, and $gt=8000$, with the initial state $|00\rangle|00\rangle_{c}$.}
\end{figure}
So far, we have discussed how to prepare the MDMS on the basis of a perfect phase matching condition $\varphi_{1(2)}=\pm0.5\pi$. Nevertheless, the effect of phase mismatch
is unavoidable in experiments. Thus it is necessary to discuss the effects caused by this error. For the sake of convenience, we suppose that the Rabi frequencies of lasers related to the collective decay
operator $S_x$ are perfect, and the phase mismatch is only introduced as executing the collective decay
operator $S_y$, and thence the effective master equation of the
first model is written as:
\begin{eqnarray}\label{f1}
\dot{\rho}_{m1}&=&\frac{1}{2}\mathscr{L}_{\gamma}[\sum_{k=1}^{2}-ie^{i\delta\varphi_{1}}|0\rangle_{k}\langle1|
+ie^{i\delta\varphi_{2}}|1\rangle_{k}\langle0|]\rho\nonumber\\&&+\frac{1}{2}\mathscr{L}_{\gamma}[S_{x}]\rho,
\end{eqnarray}
where $\delta\varphi_{i}$ denotes the phase deviation from the standard value.
Fig.~\ref{p9}(a) using the effective master equation (\ref{f1}) characterizes the the effects of phase mismatch $\delta\varphi_{1(2)}$ on the fidelity of the first scenario from the perspective of dynamic evolution. Staring from the initial state $|00\rangle|0\rangle_c$, a high-fidelity MDMS is always attainable for a long time except $\delta\varphi_{1}=-\delta\varphi_{2}=\pm 0.5\pi$ (see appendix for detail). The above conclusion is further verified in Fig.~\ref{p9}(b) whose evolution is governed by full master equation by considering $\delta\varphi_{1}=-\delta\varphi_{2}$. These results, in turn, show that the condition of dissipatively preparing the MDMS   in Ref.~\cite{EPL} is not necessary. In fact, there are many combinations of collective decay operators that can realize the MDMS. For example, the target state $\rho=(|\Phi^{+}\rangle\langle\Phi^{+}|+|\Phi^{-}\rangle\langle\Phi^{-}|+|\Psi^{+}\rangle\langle\Psi^{+}|)/3$ is also the unique state of the master equation
\begin{equation}\label{new}
\dot{\rho}=\mathscr{L}_{\gamma}[S_{x}]\rho+\mathscr{L}_{\gamma}[\chi]\rho,
\end{equation}
where
\begin{equation}\label{new1}
\chi=\begin{bmatrix}
0&-i\\
1&0
\end{bmatrix}\otimes I+I\otimes\begin{bmatrix}
0&-i\\
1&0
\end{bmatrix},
\end{equation}
corresponding to $\delta\varphi_{1}=0$ and $\delta\varphi_{2}=-0.5\pi$. In this sense, the current scheme is robust against the fluctuations of the phases of classical fields, and additionally provides us a simpler method to produce the MDMS in experiment, { i.e.}, only the phase of driving field coupled to the transition $|e\rangle\leftrightarrow|1\rangle$ needs to be alternatively changed.

As for the coupled-cavity system, since we need four classical fields to individually address two atoms to achieve the collective decay operator $S_y$, the master equation including the phase mismatch of the corresponding Rabi frequencies reads
\begin{eqnarray}\label{f2}
\dot{\rho}_{m2}&=&\mathscr{L}_{\gamma}[ie^{i\delta\varphi_{1}}|1\rangle_1\langle0|
-ie^{i\delta\varphi_{2}}|0\rangle_{1}\langle1|
+ie^{i\delta\varphi_{3}}|1\rangle_2\langle0|\nonumber\\&&
-ie^{i\delta\varphi_{4}}|0\rangle_{2}\langle1| ]\rho
+\mathscr{L}_{\gamma}[S_{x}]\rho.
\end{eqnarray}
There are four independent variables $\delta\varphi_{i}$ which is  complicated to discuss. However, if $\delta\varphi_{1}$ and $\delta\varphi_{3}$ change synchronously as well as $\delta\varphi_{2}$ and $\delta\varphi_{4}$, we can recover the result of Eq.~(\ref{f1}). But if $\delta\varphi_{1}=\delta\varphi_{2}$, and $\delta\varphi_{3}=\delta\varphi_{4}$, the uniqueness of the target state is destroyed and makes the scheme sensitive to the mismatch of the phases of classical fields, as shown in Fig.~\ref{p9}(c) and Fig.~\ref{p9}(d). Therefore, in this case the phase mismatch $|\delta\varphi|$ should be restricted below $0.1\pi$ to promise a high fidelity over $0.99$. In other cases, as long as the synchronization of $\delta\varphi_{1}$ and $\delta\varphi_{3}$ ($\delta\varphi_{2}$ and $\delta\varphi_{4}$) is ruined, the uniqueness of the system's steady state will also be destroyed.

\section{Summary}\label{VI}
In summary, our work has provided two schemes to dissipatively produce the maximally discordant mixed state where the environment becomes a resource for state generation and breaks the time limit of the unitary dynamics. In the first scheme, by alternatively changing the phase of classical fields, the target state turns into the unique steady state of the whole process, while the second one leaves out the alternating evolutionary process by introducing a lossy coupled-cavity system. We have made a comparison between two schemes. Both of them have advantages and disadvantages. For the first one, it takes shorter time to achieve the target state with the fidelity oscillated around a certain value. For the second one, although it takes a longer time to achieve the target state, the fidelity is more stable and higher. Meanwhile, both systems have favorable resistance to the spontaneous emission of atoms, and the target state can be obtained with arbitrary initial state except for the singlet state $|\Psi^{-}\rangle$. In addition, we have talked over the effect of phase mismatch on the proposed schemes and provide more mathematical forms of the master equation to prepare the MDMS. We have also discussed the relevant parameters under current experimental data and obtain high fidelities over $99\%$. We hope the work may be useful for the experimental realization on quantum correlation in the near future.

\section{Acknowledgements}
The author thank the anonymous reviewers for constructive
comments that helped in improving the quality of this paper. This work is supported by National Natural Science Foundation of China (NSFC) under Grants No. 11774047.

\appendix*
\section{THE STEADY STATE SOLUTION OF EQ.~(\ref{f1})}
In order to find the stationary solution of Eq.~(\ref{f1}), we are encouraged to expand the density operator $\rho$ in a subspace spanned by
\begin{eqnarray}\label{Ap1}
|1\rangle&=&|00\rangle,\  |2\rangle=\frac{1}{\sqrt{2}}(|01\rangle+|10\rangle),\
|3\rangle=|11\rangle.
\end{eqnarray}
Then the effective master equation (\ref{f1}) is changed into
\begin{eqnarray}\label{Ap3}
\dot{\rho}&=&\frac{1}{2}\mathscr{L}_{\gamma}[\frac{1}{\sqrt{2}}(-ie^{i\delta\varphi_{1}}|1\rangle\langle2|
+ie^{i\delta\varphi_{2}}|2\rangle\langle1|-ie^{i\delta\varphi_{2}}|2\rangle\langle3|\rho\nonumber\\&&+ie^{i\delta\varphi_{1}}
|3\rangle\langle2|)]\rho+\frac{1}{2}\mathscr{L}_{\gamma}
[\frac{1}{\sqrt{2}}(|1\rangle\langle2|+|2\rangle\langle3|)\nonumber\\&&+\mathrm{H.c.}]\rho,
\end{eqnarray}
the steady state solution of the above equation can be solved by $\dot{\rho}=0$ and we have
\begin{widetext}
\begin{align}\label{Ap2}
\begin{bmatrix}\begin{smallmatrix}
[e^{-i\delta}\rho_{13}+e^{i\delta}\rho_{31}
-(4\rho_{11}+\rho_{13}-4\rho_{22}+\rho_{31})]\gamma&
[e^{i\delta}(-2\rho_{21}+\rho_{32})-(6\rho_{12}-2\rho_{21}-4\rho_{23}+\rho_{32})]\gamma&
[3\rho_{22}-4\rho_{13}-1-e^{i\delta}(3\rho_{22}-1)]\gamma\\
[e^{-i\delta}(\rho_{23}-2\rho_{12})+2\rho_{12}-6\rho_{21}-\rho_{23}+4\rho_{32}]\gamma&
-2(e^{-i\delta}\rho_{13}+e^{i\delta}\rho_{31}
+6\rho_{22}-\rho_{31}-\rho_{13}-2)\gamma&
[e^{i\delta}(\rho_{21}-2\rho_{32})+4\rho_{12}+2\rho_{32}-\rho_{21}-6\rho_{23}]
\gamma\\
[e^{-i\delta}(1-3\rho_{22})+3\rho_{22}-4\rho_{31}-1]\gamma&
[e^{-i\delta}(\rho_{12}-2\rho_{23})-\rho_{12}+4\rho_{21}+2\rho_{23}
-6\rho_{32}]\gamma&
(e^{-i\delta}\rho_{13}+e^{i\delta}\rho_{31}
+4\rho_{11}+8\rho_{22}-\rho_{13}-\rho_{31}-4)\gamma
\end{smallmatrix}\end{bmatrix}=0,
\end{align}
\end{widetext}
where $\delta=\delta\varphi_{1}-\delta\varphi_{2}$. It can be examined both numerically and analytically that Eq.~(\ref{Ap2}) has a unique steady state solution except $\delta\varphi_{1}=-\delta\varphi_{2}=\pm 0.5\pi$.
\bibliography{pra}

\end{document}